\documentclass[conference]{IEEEtran}
\IEEEoverridecommandlockouts
\usepackage{cite}
\usepackage[numbers]{natbib}
\usepackage{amsmath,amssymb,amsfonts}
\usepackage{algorithmic}
\usepackage{graphicx}
\usepackage{flushend}
\usepackage{dblfloatfix}
\usepackage{textcomp}
\usepackage{xcolor}
\usepackage{url}
\usepackage{tcolorbox}
\usepackage{subfigure}
\usepackage{multirow}
\usepackage{ulem}
\usepackage{balance}

 \usepackage[colorlinks = true,
            linkcolor = blue,
            urlcolor  = blue,
            citecolor = black,
            anchorcolor = blue]{hyperref}
            
\usepackage{colortbl}
\usepackage{tikz}
\usetikzlibrary{plotmarks}
\usetikzlibrary{arrows,shapes,positioning}
\usetikzlibrary{decorations.markings}
\tikzstyle arrowstyle=[scale=1]
\tikzset{>=latex}

\newcommand{\sub}{\textit{pre}-review}
\newcommand{\rev}{\textit{post}-review}

\newcommand{\PqOne}{$PQ_1$}
\newcommand{\PqTwo}{$PQ_2$}

\newcommand{\RqOne}{{$RQ_1$}}
\newcommand{\RqTwo}{$RQ_2$}

\newcommand{\PqOneSen}{\PqOne~What kinds of files churn during a review?}

\newcommand{\PqTwoSen}{\PqTwo~Which kinds of reviewed files are likely to conform after the review process?}

\newcommand{\RqOneSen}{\RqOne~Review Process: Does a patch coding pattern conform to the project after it has been reviewed?}
\newcommand{\RqTwoSen}{\RqTwo~Patch Decision: Are there coding pattern differences between an accepted and an abandoned patch?}

\def\BibTeX{{\rm B\kern-.05em{\sc i\kern-.025em b}\kern-.08em
    T\kern-.1667em\lower.7ex\hbox{E}\kern-.125emX}}
    
\begin{document}

\title{Does Code Review Promote Conformance? \\ A Study of OpenStack Patches}

\author{
\IEEEauthorblockN{
    Panyawut Sri-iesaranusorn, 
    Raula Gaikovina Kula, 
    Takashi Ishio
    }
\IEEEauthorblockA{
    {Nara Institute of Science and Technology, Nara, Japan}\\
    {Email: \{sri-iesaranusorn.panyawut.sg0, raula-k, ishio\}@is.naist.jp}\\
    }
}
\maketitle

\begin{abstract}
Code Review plays a crucial role in software quality, by allowing reviewers to discuss and critique any new patches before they can be successfully integrated into the project code.
Yet, it is unsure the extent to which coding pattern changes (i.e., repetitive code) from when a patch is first submitted and when the decision is made (i.e., during the review process).
In this study, we revisit coding patterns in code reviews, aiming to analyze whether or not the coding pattern changes during the review process.
Comparing prior submitted patches, we measure differences in coding pattern between {\sub}~(i.e., patch before the review) and {\rev}~(i.e., patch after a review) from 27,736 reviewed OpenStack patches.
Results show that patches after review, tend to conform to similar coding patterns of accepted patches, compared to when they were first submitted.
We also find that accepted patches do have similar coding patterns to prior accepted patches.
Our study reveals insights into the review process, supporting the potential for automated tool support for newcomers and lays the groundwork for work into understanding conformance and how it makes for an efficient code review process.
\end{abstract}

\section{Introduction}
Code Review plays a crucial role in software quality by allowing reviewers to discuss and critique new patches before they can be successfully integrated into the project code.
As a code quality assurance activity, code review also functions as knowledge transfer, team building, and coordination mechanism within software teams~\cite{Bacchelli_2013}.
It has approval from industry giants like Microsoft and Google, on how `Code Reviews at Microsoft are an integral part of the development process that thousands of engineers perceive it as a great best practice and most high-performing teams spend a lot of time doing'. 
Although light-weight variations of code review streamline the process, reviews still suffer from being ineffective and less efficient.
This can detract future contributions, especially for Open Source Software projects that need to attract and retain newcomer contributions.

Prior work has shown that similar coding patterns (i.e., \textit{such as the way developers edit day-to-day code that tends to be repetitive, often using existing code elements}) is related to the acceptance of a patch.
\citet{Hellendoorn2015} shows that reviewers may consider conformance to the project’s code style as an indicator of acceptance. 
Comparing the submitted code with code in the project by using language models \cite{robbesASE2010}, they found that rejected changesets contain code significantly less similar to the project.
Thus, it is not known the extent to which coding pattern changes during the review process, from when a patch is first submitted to when it is accepted.

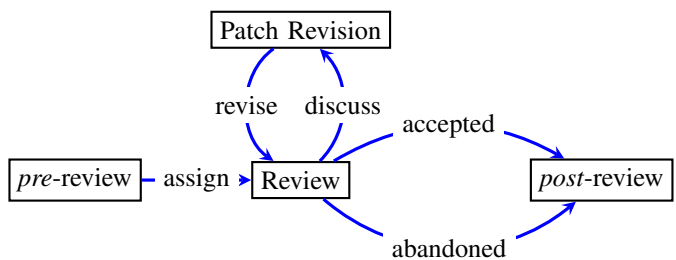
\begin{figure}[t]
	\centering
	\begin{tikzpicture}
	\begin{scope}[every node/.style={thick,draw}]
	\node (r) at (-3,2) {\sub};
	\node (a) at (0,2) {Review};
	\node (rr) at (0,4) {Patch Revision};
	\node (d) at (4,2) {\rev};
	\end{scope}
	\begin{scope}[>={stealth},
	every node/.style={fill=white},
	every edge/.style={draw=blue,very thick}]
	\path [->] (r) edge[bend right=0 ] node {assign} (a); 
	\path [->] (a) edge[bend right=45] node {discuss} (rr); 
	\path [->] (rr) edge[bend right=55] node {revise} (a); 
	\path [->] (a) edge[bend right=40] node {abandoned} (d);
	\path [->] (a) edge[bend right=-30] node {accepted} (d); 
	\end{scope}
	\end{tikzpicture}
	\caption{The Code Review process shows key activities between the {\sub}~and {\rev}~of a submitted patch.}
	\label{fig:relationships}
	\vspace{-0.3cm}
\end{figure}

To fill this gap, in this study, we revisit coding patterns in code reviews, aiming to analyze whether or not the coding pattern changes during the review process.
We define the conformance as "\textit{similarly repetitive written code patterns that appear in prior accepted patches for a project}".
We reuse language models to compare coding patterns or prior submitted patches before the review (i.e., \sub) with a revised patch after the review (i.e., \rev), as shown in Figure \ref{fig:relationships}.
Also different to \citep{Hellendoorn2015}, we study patches, which are atomic, modular, and updated by design as opposed to pull requests, that are more a collection of commits, changed files, and the differences (or "diff") between files in branches \cite{ReviewProposed}.

Our large-scale empirical study consisted of two parts.
First, we conduct a preliminary study of 27,736 reviewed patches taken from the OpenStack projects.
The preliminary results show that the programming files tend to contain more churn and are likely to contain coding pattern changes when compared to configuration or documentation types of files.
Focusing on the programming files (i.e., JavaScript, Bash Shell, and Python), we then form two research questions to guide our study.

\begin{itemize}
    \item \textit{{\RqOneSen}}
    \item \textit{{\RqTwoSen}}
\end{itemize}

\begin{figure*}[]
    \begin{center}
    \centerline{\includegraphics[width=\textwidth]{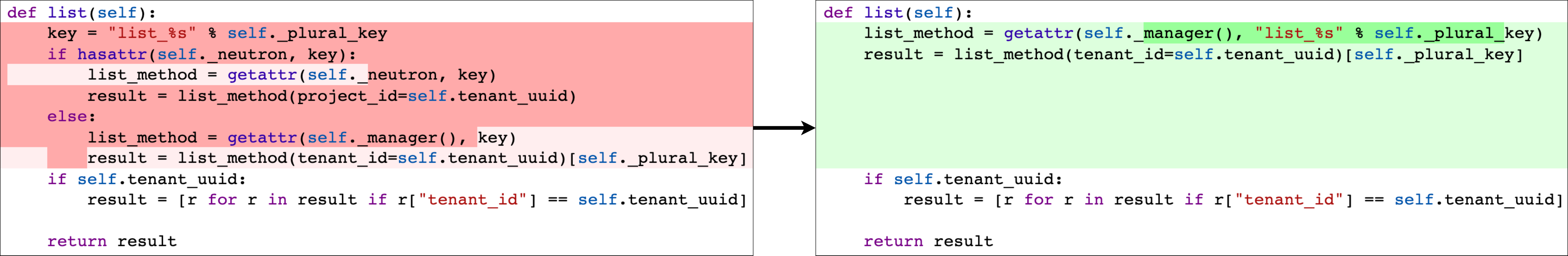}}
    \caption{An example of coding pattern changes between \sub~and \rev. We can see that \rev~code has changed its coding pattern to change the \texttt{key} variable into the \texttt{list\_method} variable. The code excluding \texttt{if}-\texttt{else} statement is consistent with other patches that have been accepted.}
    \label{fig:motivation}
    \end{center}
\end{figure*}

For RQ$_1$, results show that patches after being reviewed, tend to conform to similar coding patterns of accepted patches, compared to when they were first submitted.
The results of RQ$_2$ also show that the conformance of accepted patches is higher than patches that were abandoned, which confirms prior findings.
Our full replication package can be found at \url{https://zenodo.org/record/4537542#.YCaArhMza3I}

\section{Motivating Example: Identifying Similar Coding Patterns using a language model.}

Figure \ref{fig:motivation} shows a review example \cite{opendev} on how we compare the coding patterns that appear in the \rev~ compared to the \sub.
In this example, we see that the patch purpose is to add a helper function that \textit{"is a way to unify the way of the usage of the Neutron API and remove code duplication"}.
This patch has been through 50 patch revisions and consists of 14 python files.
In over fifty patch revisions of the patch, we see that evidence of coding pattern changes with the variable \texttt{key} being changed into \texttt{list\_method} variable.
If the \texttt{list\_method} variable is consistent in other prior accepted patches, we regard this as conformance to such repetitive coding patterns in the patch.

The statistical counts infer the probability of conformance to the coding pattern as the entropy.
Low cross-entropy indicates the high similar style or conformance. 

\begin{table}[b]
\centering
\caption{Corpus size of tokens for each file extension.}
\label{tab:corpus}
\begin{tabular}{crrrr}
\hline
\multicolumn{1}{l}{\textbf{File ext.}} & \multicolumn{1}{c}{\textbf{\#Revisions}} & \multicolumn{1}{c}{\textbf{\#Files}} & \multicolumn{1}{l}{\textbf{\#Unique Token}} & \multicolumn{1}{c}{\textbf{\#Token}} \\ \hline
.py & 18,859 & 112,566 & 513,280 & 816,364,358 \\
.php & 20 & 21,723 & 11,535 & 5,330,917 \\
.yaml & 4,575 & 12,295 & 26,301 & 79,673,298 \\
.rst & 4,560 & 6,761 & 38,372 & 12,251,290 \\
.pp & 940 & 4,974 & 15,301 & 4,858,510 \\
.yml & 960 & 3,976 & 10,811 & 1,636,665 \\
.sh & 2,307 & 3,946 & 17,482 & 8,255,596 \\
.json & 1,137 & 2,977 & 20,563 & 6,275,269 \\
.txt & 2,105 & 2,961 & 2,283 & 474,107 \\
.js & 567 & 2,542 & 29,575 & 11,478,552 \\
.html & 512 & 1,772 & 6,588 & 703,987 \\
.xml & 253 & 1,333 & 4,386 & 597,631 \\
\hline
\end{tabular}
\end{table}

\section{Data Preparation}
\label{sec:exp}

For our dataset collection, we acquired OpenStack patches from \citet{UedaIWESEP18}, which is patch source code originally mined from the Gerrit API.
The detail of each step is explained in the following steps:
\textit{Step1: Label pre and post reviews -} 
    Our data was initially in JSON format, which is annotated as added, removed, or unchanged lines. 
    From this data we create {\sub} and {\rev}, containing the appropriate lines -- added lines to only {\rev}, removed to only {\sub}, and unchanged to both. 
\textit{Step2: Remove comments - }
    To remove the comments, we use the NCDSearch tool \cite{Ishio2018}.
    The tool applies grammar from the lexer files generated by ANTLR4 parser generator \cite{ANTLR}.
\textit{Step3: Tokenize source code - } 
    In addition to removing comments, we also used NCDSearch as our tokenization tool as it supports several languages such as Python, JavaScript, and plain text like .txt and .html. 

For training a language model, we use the MITLM toolkit implemented in \cite{MITLM}.
To understand how style changes due to the parameter $n$ of the n-gram models, we measure the model created from 3-grams to 9-grams. 

\section{Coding Patterns of File Types}
\label{sec:preliminary result}

To address this gap, we perform a file-level analysis focused on the following two Preliminary Questions (PQs):
\begin{itemize}
    \item \textit{\PqOneSen} 
    We want to study the number of patch revisions of a typical review to better understand the magnitude of patches submitted.
    \item \textit{\PqTwoSen} There is no prior work that quantifies which type of files tend to confirm after review. 
\end{itemize}

Table \ref{tab:corpus} shows statistics of the collected corpus that belongs to the OpenStack project.
This dataset will be used for all the preliminary questions and contains 27,736 revision patches over 177,826 files.
Both the {\sub} and {\rev} combined contain 900M tokens with 0.70M unique tokens. 

To evaluate the kinds of files, we use the same groupings employed by \citet{Mcintosh_2014}.
Hence, we classify each unique file that ever existed in the analyzed time span as either configuration, programming, or documentation files. 
The list below shows the file extensions that we manually investigated and classified:
\begin{itemize}
    \item \textit{Configuration files (conf.)}: The file extensions include \texttt{.yml, .json, .xml, .pp, and .yaml}.
    \item \textit{Programming files (prog.)}: The file extensions include \texttt{.sh, .py,  and .js}.
    \item \textit{Documentation (doc.)}: The file extensions include \texttt{.rst, .php, .html, and .txt}.
\end{itemize}

\begin{figure}[]
    \centerline{\includegraphics[width=.8\columnwidth]{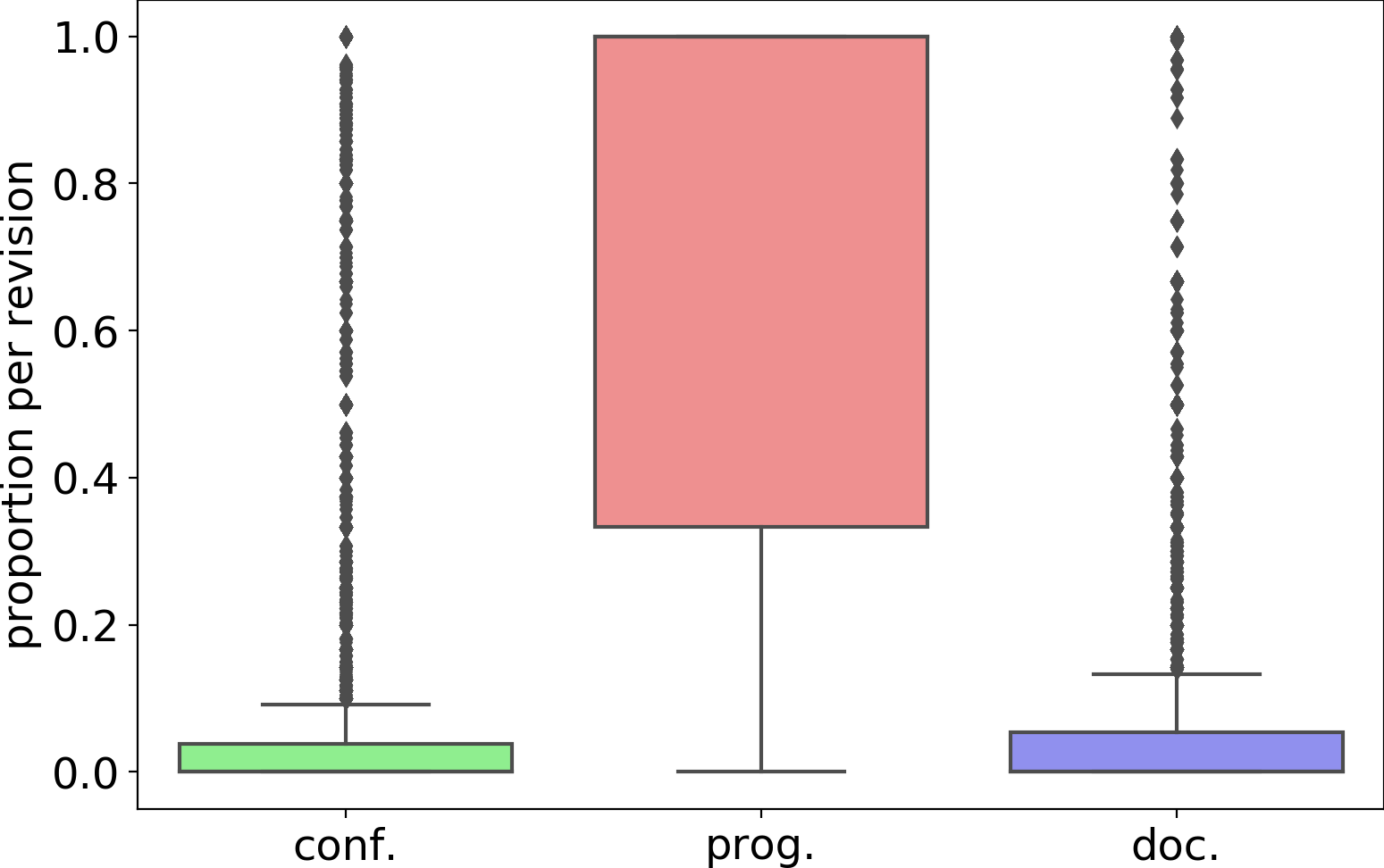}}
    \caption{File churn for each kind of files during a review. Programming file type have more churn compared to configuration and documentation types.}
    \label{fig:PQ1}
\end{figure}

Figure \ref{fig:PQ1} shows the results that answer to \PqOne, confirming that the programming files tend to churn more than other kinds of files. 
As shown in the figure, the average proportions of programming, configuration, and documentation are 69.97\%, 14.15\%, and 15.88\%.
The most common files are the source code files for Bash shell scripts, Python, and JavaScript files.
We test the null hypothesis that \textit{``the file churn for all different kinds of files is the same''}.
Our result shows that there are statistically significant differences between two or more groups in our data $(p < .001)$.
This preliminary result suggests that during a code review, the intuitive focus is on code inspection, thus programming files will receive more changes.

\begin{figure}[]
    \centerline{\includegraphics[width=.8\columnwidth]{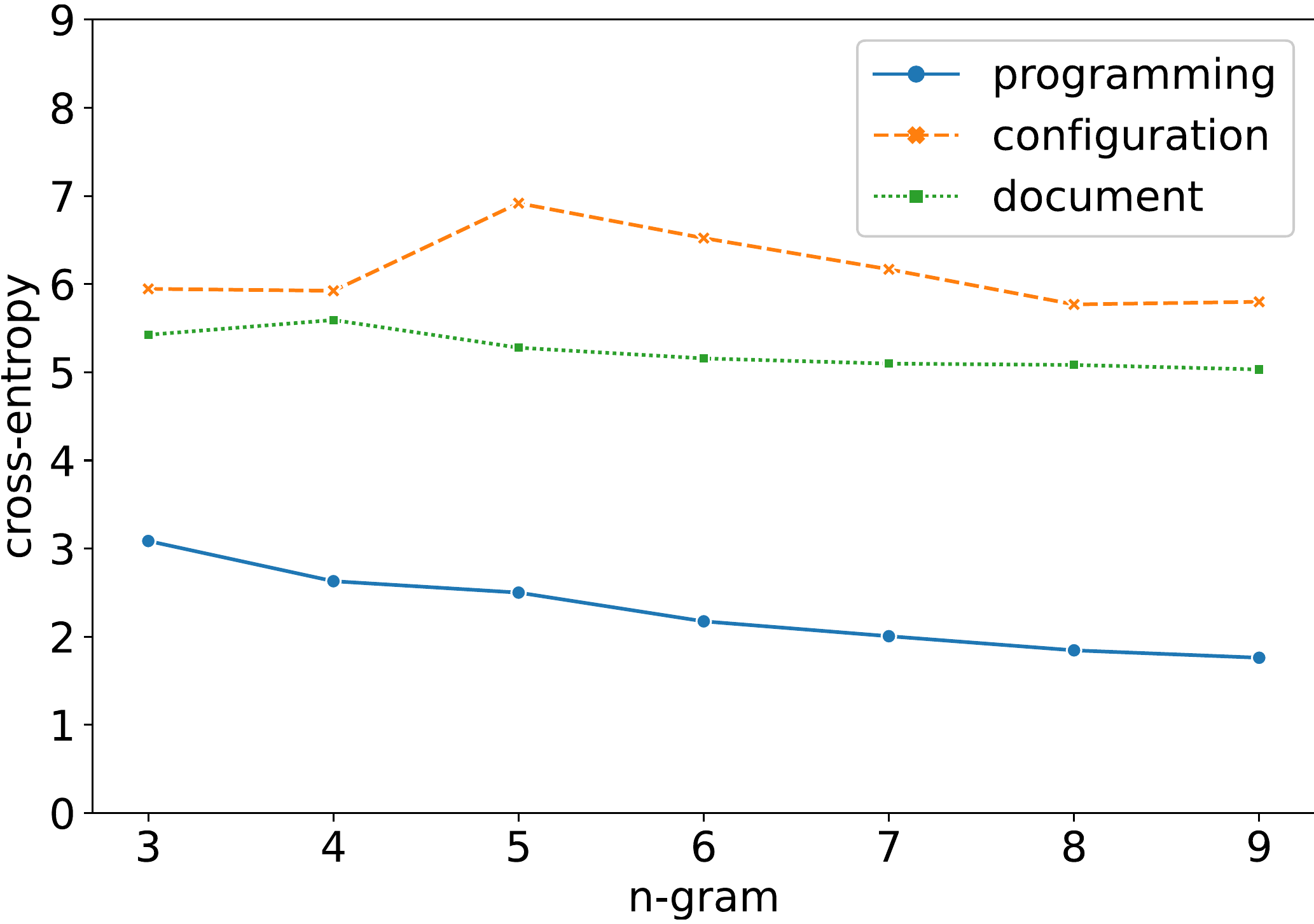}}
    \caption{Analysis of coding pattern changes (as n-grams) between the different types of files during a review.}
    \label{fig:PQ2}
\end{figure}

Figure \ref{fig:PQ2} shows the results relating to \PqTwo, where the programming related files are more likely to be repetitive.
The entropy of programming file type is 3.086, and 1.762 for 3-grams and 9-grams, respectively.
For configuration and documentation/other groups, the entropy changed from 5.946, and 5.425 to 5.799, and 5.031 for 3-grams and 9-grams.
Considering the results of each group, the entropy of the programming decreased when $n$ increases, while the entropy of other groups seems increased or stable.
One reason for a higher entropy is that natural language may use different terms that vary across patches that may have different functions and descriptions.
As a result, these two groups are less conformant than the programming group, which are constrained by the syntax of programming languages. 
Therefore, we focused on the programming group for our experiment.

\begin{tcolorbox}
    \textbf{Summary}:
    Programming files (code) tend to contain more churn, and are more likely to conform to the coding pattern compared to configuration or documentation files.
\end{tcolorbox}

\begin{table}[]
\centering
\caption{Percentage of language syntax token.}
\label{tab:syntax}
\begin{tabular}{lccc}
\hline
\multicolumn{1}{l}{} & \multicolumn{1}{c}{\textbf{Separators}} & \multicolumn{1}{c}{\textbf{Operators}} & \textbf{Keywords} \\ \hline \hline
\multicolumn{4}{l}{\textbf{.py files}} \\ 
\multicolumn{1}{l}{Pre-review} & \multicolumn{1}{c}{33.88\%} & \multicolumn{1}{c}{5.64\%} & 4.81\% \\ 
\multicolumn{1}{l}{Post-review} & \multicolumn{1}{c}{33.97\%} & \multicolumn{1}{c}{5.65\%} & 4.79\% \\ \hline
\multicolumn{4}{l}{\textbf{.js files}} \\
\multicolumn{1}{l}{Pre-review} & \multicolumn{1}{c}{44.37\%} & \multicolumn{1}{c}{8.40\%} & 6.50\% \\ 
\multicolumn{1}{l}{Post-review} & \multicolumn{1}{c}{43.83\%} & \multicolumn{1}{c}{8.09\%} & 6.47\% \\ \hline
\multicolumn{4}{l}{\textbf{.sh files}} \\ 
\multicolumn{1}{l}{Pre-review} & \multicolumn{1}{c}{-} & \multicolumn{1}{c}{5.67\%} & 3.01\% \\ 
\multicolumn{1}{l}{Post-review} & \multicolumn{1}{c}{-} & \multicolumn{1}{c}{5.69\%} & 2.99\% \\ \hline
\multicolumn{4}{l}{\textbf{Reported Statistics from Rahman et al. \cite{Rahman:2019}}} \\ 
\multicolumn{1}{l}{py} & \multicolumn{1}{c}{{41.98\%}} & \multicolumn{1}{c}{6.42\%} & 4.99\% \\ 
\multicolumn{1}{l}{js} & \multicolumn{1}{c}{47.21\%} & \multicolumn{1}{c}{6.53\%} & 6.87\% \\ \hline
\end{tabular}
\end{table}

\section{RQ$_1$: Review Process}
\label{sec:RQ1}

\begin{figure}[]
    \begin{center}
    \centerline{\includegraphics[width=.8\columnwidth]{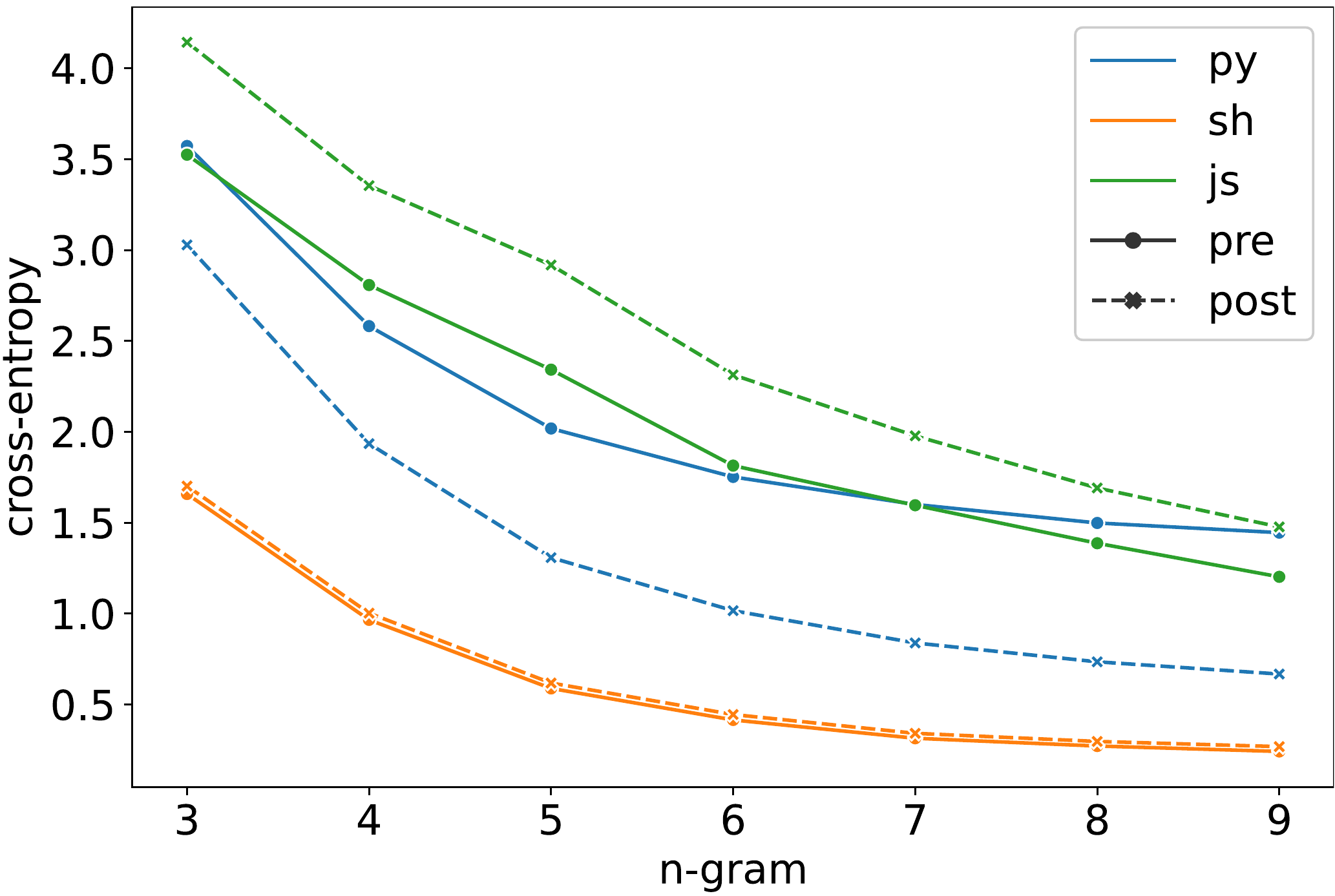}}
    \caption{{\sub} vs. {\rev}~versions based on programming type.}
    \label{fig:PrePost}
    \end{center}
\end{figure}

Figure \ref{fig:PrePost} shows the changed entropy of {\rev} compared to {\sub}, that answer to \RqOne, where the patch coding pattern conforms to the project after it has been reviewed.
A general trend can be seen that the entropy of the three file types decreases as the length of the n-gram model increases, which follows the typical trend between n-gram entropy and its structural language. 
Looking at each graph individually in Figure \ref{fig:PrePost}, there is the typical natural trend in entropy of the {\sub} and {\rev}~patches of each file type.
When we compare the two graphs, we see that {\rev}~patches tend to have a lower entropy than {\sub}. 
This is further highlighted by the python files, which tend to have lower entropy after the {\rev}. 
The result suggests that the python programmer concerns the conformance during a code review more than others.

\begin{table*}
\centering
\caption{Top 3 changed syntax tokens (added and removed) for each file type where the green and red color represent unchanged and changed token, respectively, during a code review.}
\label{tab:top3}
\scalebox{0.85}{
\begin{tabular}{ccccccc|cccccc} 
\hline
\multicolumn{7}{c|}{\textbf{Added Syntax} } & \multicolumn{6}{c}{\textbf{Removed Syntax} } \\
\multirow{2}{*}{\textbf{Top} } & \multicolumn{2}{c}{\textbf{Separator} } & \multicolumn{2}{c}{\textbf{Operator} } & \multicolumn{2}{c|}{\textbf{Keyword} } & \multicolumn{2}{c}{\textbf{Separator} } & \multicolumn{2}{c}{\textbf{Operator} } & \multicolumn{2}{c}{\textbf{Keyword} } \\
 & \textbf{token } & \textbf{percent } & \textbf{token } & \textbf{percent } & \textbf{token } & \textbf{percent } & \textbf{token } & \textbf{percent } & \textbf{token } & \textbf{percent } & \textbf{token } & \textbf{percent } \\ 
\hline \hline
\multicolumn{13}{l}{\textbf{.py files }} \\
1 & {\cellcolor[rgb]{0.78,1,0.784}}. & 20.38\% & {\cellcolor[rgb]{0.78,1,0.784}}= & 8.33\% & {\cellcolor[rgb]{0.78,1,0.784}}def & 1.30\% & {\cellcolor[rgb]{0.78,1,0.784}}. & 20.18\% & {\cellcolor[rgb]{0.78,1,0.784}}= & 8.47\% & {\cellcolor[rgb]{0.78,1,0.784}}def & 1.40\% \\
2 & {\cellcolor[rgb]{0.78,1,0.784}}, & 13.19\% & {\cellcolor[rgb]{0.78,1,0.784}}in & 0.48\% & {\cellcolor[rgb]{0.78,1,0.784}}if & 0.84\% & {\cellcolor[rgb]{0.78,1,0.784}}, & 13.26\% & {\cellcolor[rgb]{0.78,1,0.784}}in & 0.59\% & {\cellcolor[rgb]{0.78,1,0.784}}if & 0.86\% \\
3 & {\cellcolor[rgb]{0.78,1,0.784}}) & 12.99\% & {\cellcolor[rgb]{1,0.749,0.749}}- & 0.31\% & {\cellcolor[rgb]{1,0.749,0.749}}return & 0.48\% & {\cellcolor[rgb]{0.78,1,0.784}}) & 12.23\% & {\cellcolor[rgb]{1,0.749,0.749}}\% & 0.26\% & {\cellcolor[rgb]{1,0.749,0.749}}in & 0.59\% \\ 
\hline
\multicolumn{13}{l}{\textbf{.js files }} \\
1 & {\cellcolor[rgb]{1,0.749,0.749}}. & 15.02\% & {\cellcolor[rgb]{0.78,1,0.784}}= & 7.34\% & {\cellcolor[rgb]{1,0.749,0.749}}this & 3.31\% & {\cellcolor[rgb]{1,0.749,0.749}}( & 14.73\% & {\cellcolor[rgb]{0.78,1,0.784}}= & 6.49\% & {\cellcolor[rgb]{1,0.749,0.749}}var & 2.66\% \\
2 & {\cellcolor[rgb]{1,0.749,0.749}}( & 13.89\% & {\cellcolor[rgb]{1,0.749,0.749}}\textless & 1.80\% & {\cellcolor[rgb]{1,0.749,0.749}}var & 2.58\% & {\cellcolor[rgb]{1,0.749,0.749}}. & 14.50\% & {\cellcolor[rgb]{1,0.749,0.749}}+ & 1.00\% & {\cellcolor[rgb]{1,0.749,0.749}}this & 2.55\% \\
3 & {\cellcolor[rgb]{0.78,1,0.784}}) & 13.47\% & {\cellcolor[rgb]{1,0.749,0.749}}\textgreater & 1.69\% & {\cellcolor[rgb]{0.78,1,0.784}}return & 1.76\% & {\cellcolor[rgb]{0.78,1,0.784}}) & 14.48\% & {\cellcolor[rgb]{1,0.749,0.749}}\textless & 0.91\% & {\cellcolor[rgb]{0.78,1,0.784}}return & 1.88\% \\ 
\hline
\multicolumn{13}{l}{\textbf{.sh files }} \\
1 &  &  & {\cellcolor[rgb]{0.78,1,0.784}}- & 39.31\% & {\cellcolor[rgb]{1,0.749,0.749}}\{ & 2.59\% &  &  & {\cellcolor[rgb]{0.78,1,0.784}}- & 32.57\% & {\cellcolor[rgb]{1,0.749,0.749}}then & 3.47\% \\
2 &  &  & {\cellcolor[rgb]{0.78,1,0.784}}/ & 30.49\% & {\cellcolor[rgb]{1,0.749,0.749}}\} & 2.28\% &  &  & {\cellcolor[rgb]{0.78,1,0.784}}/ & 23.62\% & {\cellcolor[rgb]{1,0.749,0.749}}\{ & 3.17\% \\
3 &  &  & {\cellcolor[rgb]{0.78,1,0.784}}= & 5.67\% & {\cellcolor[rgb]{1,0.749,0.749}}then & 2.28\% &  &  & {\cellcolor[rgb]{0.78,1,0.784}}= & 9.87\% & {\cellcolor[rgb]{1,0.749,0.749}}if & 3.17\% \\
\hline
\end{tabular}
}
\end{table*}

Table \ref{tab:syntax} serves as a sanity check to compare our work to prior work in terms of the separators, operators, and keywords proportions.
Although the percentages of separators for python files have a 10\% difference to the \citet{Rahman:2019} study, we find that the {\sub} and {\rev} percentages themselves are fairly consistent.
Moreover, as shown in Table \ref{tab:top3}, we conclude that there is no clear coding pattern between the separator, operator, and the keyword during a code review.
One possible reason is that only a percentage of these tokens maybe not enough to find the coding pattern.

A key takeaway message is patches tend to contain natural coding pattern and more conformance after review, particularly for python files.
Potential future work is to manually investigate whether the natural coding pattern correlates with complexity \rev.
Similar to \citet{Campbell_2014}, a syntax suggestion or highlighter tool during a review is feasible.

\begin{tcolorbox}
    \textbf{Summary}: 
    We provide evidence that the review process changes the coding pattern of the patch.
    Results show that the conformance of a patch after being reviewed, tend to be higher than a patch that was first submitted.
\end{tcolorbox}

\section{RQ$_2$: Patch Decision}
\label{sec:RQ2}

\begin{figure}[]
    \begin{center}
    \centerline{\includegraphics[width=.8\columnwidth]{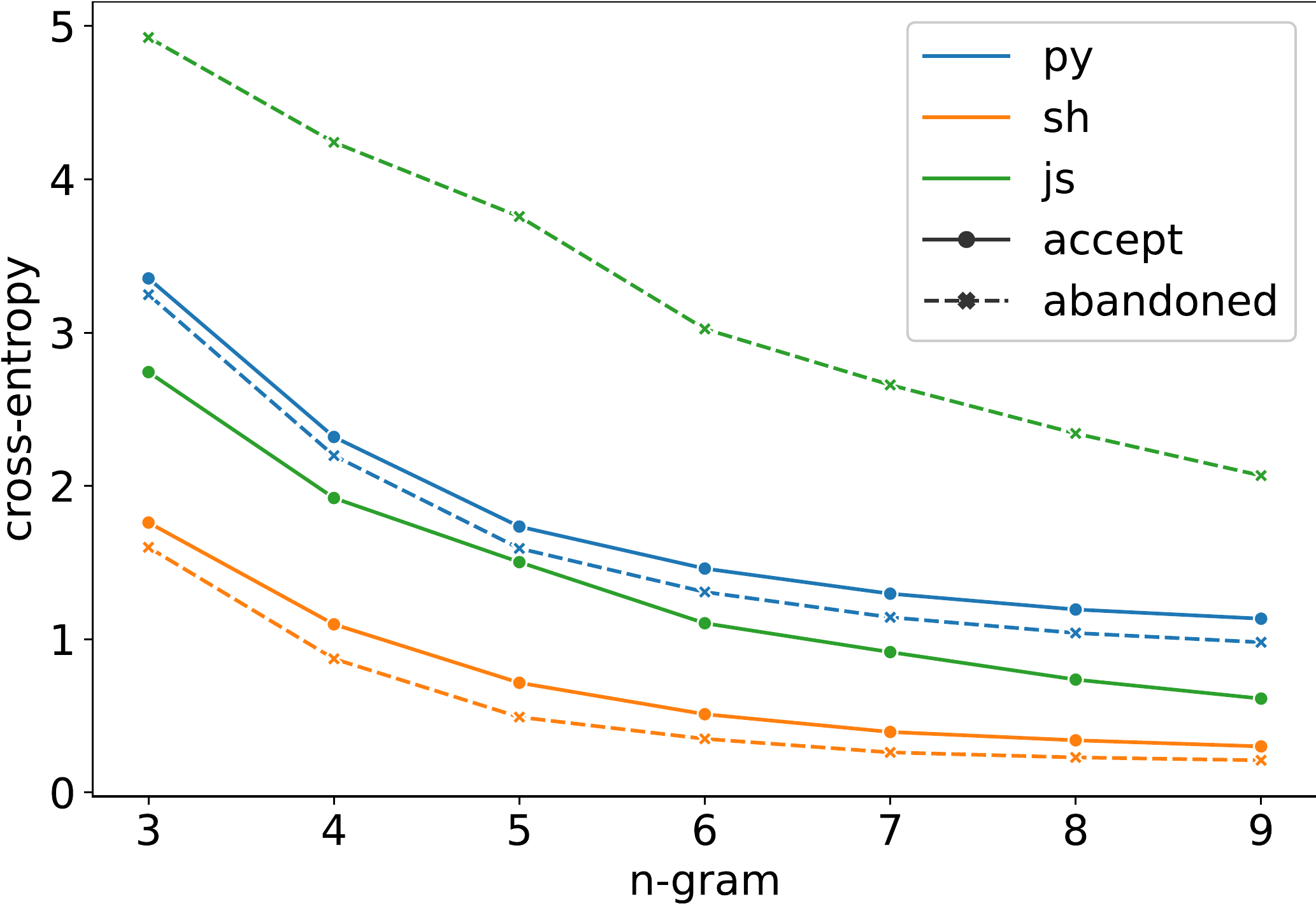}}
    \caption{Accepted vs. Abandoned Patches on programming language file type. The green line represents the .js file type that its accepted patches are more likely to conform when compared to the abandoned patches.}
    \label{fig:AcceptAbandon}
    \end{center}
\end{figure}

Figure \ref{fig:AcceptAbandon} shows the cross-entropy of each programming file type based on whether it was accepted or not.
Similar to RQ$_1$, the entropy trend and structural language are similar.
It suggest that there is a difference in the coding pattern between accepted and abandoned patch groups. 
Confirming the related work of \cite{Hellendoorn2015}, we see that accepted patches are more conformant than the abandoned patches.
At the file-level, the entropy of JavaScript file type is lower than Python file type in accepted patches, which is not the case for abandoned patches.
This leads us to suspect that JavaScript developers concern the unknown factor which possibly is more important than conformance in the first stage of implementation.

One important takeaway from the results of $RQ_2$ is that patches themselves are sufficient to measure the coding pattern of the patch.
This is instead of compared to analyzing the code in the project itself,
which is performed by prior work.
Analysis of the patches themselves should pave the way for potential code support tools and recommendation tools.
Similar to $RQ_1$, a syntax suggestion or highlighter tool during a review is feasible, but for this case, we would like to predict the likelihood that a patch can be accepted or not.

\begin{tcolorbox}
    \textbf{Summary}: 
    The results suggest that there are differences in the coding pattern between accepted and abandoned patches groups.
    This is especially the case for the JavaScript code of the patch.
\end{tcolorbox}

\section{Summary and Future Challenges}
\label{sec:conclusion}
Our preliminary study confirms that reviewing code conforms code to repetitive coding patterns.
Researchers and OSS project teams could use our results as motivation for exploring tool support or automatic detection of conformance coding patterns, while newcomers could increase the likelihood for acceptance by looking at prior submitted patches.

This work lays the groundwork to open new avenues such as exploring (1) whether or not there is a universal definition of conformance for source code and (2) whether conformance is related to developer individual experience or skill, and (3) whether or not conformance creates a more efficient review process, to name a few.
Furthermore, we plan to explore how our approach contrasts and complements other techniques such as  coding style checkers \cite{Han_2020}.  

\section*{Acknowledgement}
This work supported by JSPS KAKENHI Grants JP20H05706, JP18H03221, JP20K19774, and JP18H04094.

\bibliographystyle{plainnat}
\newpage
\bibliography{ownrefs}
\flushend
\end{document}